\documentclass[sigconf,screen,sigconf]{acmart}
\copyrightyear{2025}
\acmYear{2025}
\setcopyright{rightsretained}
\acmConference[CF Companion '25]{22nd ACM International Conference on Computing Frontiers}{May 28--30, 2025}{Cagliari, Italy}
\acmBooktitle{22nd ACM International Conference on Computing Frontiers (CF Companion '25), May 28--30, 2025, Cagliari, Italy}\acmDOI{10.1145/3706594.3726981}
\acmISBN{979-8-4007-1393-4/2025/05}

\usepackage{acmart-taps}
\usepackage{cleveref}
\usepackage[
    per-mode=symbol,
    per-symbol =/,
    separate-uncertainty=true,
    multi-part-units=single
]{siunitx}
\usepackage[acronym,nohypertypes={acronym}]{glossaries}

\usepackage{multirow}
\usepackage{subcaption}
\usepackage{enumitem}
\usepackage{soul}

\newcommand{\RedMulE}{\textit{RedMulE-FT}}

\newcommand{\icon}[1]{({#1})}

\DeclareSIUnit\gateequivalent{GE}

\def\cglspl#1{\glspl{#1}}

\def\cgls#1{\gls{#1}}

\makeatletter
\newif\ifauthordraft
\newif\ifnotauthordraft

\newif\ifreview
\newif\ifnotreview
\makeatother

    \newcommand{\todo}[1]{}
    \newcommand{\R}[1]{#1}
    
    \newcommand{\repolink}{\url{https://github.com/pulp-platform/redmule-ft}}

\newcommand{\resultareabaseline}{\R{\SI{583}{\kilo\gateequivalent}}}
\newcommand{\resultareadata}{\R{\SI{596}{\kilo\gateequivalent}}}
\newcommand{\resultareafull}{\R{\SI{730}{\kilo\gateequivalent}}}

\newcommand{\resultareaoverheaddata}{\R{\SI{2.3}{\percent}}}
\newcommand{\resultareaoverheadfull}{\R{+\SI{22.9}{\percent}}}
\newcommand{\resultareaoverheadtotal}{\R{\SI{25.2}{\percent}}}
\newcommand{\resultfrequency}{\R{\SI{500}{\mega\hertz}}}
\newcommand{\resultvulnerability}{\R{$11\times$}}

\newacronym{soc}{SoC}{system-on-chip}
\newacronym{isa}{ISA}{instruction set architecture}
\newacronym{pulp}{PULP}{parallel ultra-low power}
\newacronym{odrg}{ODRG}{on-demand redundancy grouping}
\newacronym{tcls}{TCLS}{triple-core lockstep}
\newacronym{ecc}{ECC}{error correction codes}
\newacronym{dut}{DUT}{device under test}
\newacronym{wdt}{WDT}{watchdog timer}
\newacronym{seu}{SEU}{single-event upset}
\newacronym{see}{SEE}{single-event effect}
\newacronym{sefi}{SEFI}{single-event functional interrupt}
\newacronym{set}{SET}{single-event transient}
\newacronym{sel}{SEL}{single-event latchup}
\newacronym{sdc}{SDC}{silent data corruption}
\newacronym{due}{DUE}{detectable unrecoverable error}
\newacronym{tid}{TID}{total ionizing dose}
\newacronym{rhbd}{RHBD}{radiation hardened by design}
\newacronym{ge}{GE}{gate equivalent}
\newacronym{udma}{$\mu$DMA}{I/O DMA}
\newacronym{gpio}{GPIO}{general purpose input/output}
\newacronym{dma}{DMA}{direct memory access}
\newacronym{tcdm}{TCDM}{tightly coupled data memory}
\newacronym{pdk}{PDK}{process design kit}
\newacronym{pcb}{PCB}{printed circuit board}
\newacronym{sram}{SRAM}{static random-access memory}
\newacronym{tmr}{TMR}{triple modular redundancy}
\newacronym{dmr}{DMR}{dual modular redundancy}
\newacronym{secded}{SECDED}{single error correction, double error detection}
\newacronym{mftf}{MFTF}{mean fluence to failure}
\newacronym{mttf}{MTTF}{mean time to failure}
\newacronym{rpi}{RPi}{Raspberry Pi}
\newacronym{ff}{FF}{flip-flop}
\newacronym{fpga}{FPGA}{field-programmable gate array}
\newacronym{fsm}{FSM}{finite-state machine}
\newacronym{fma}{FMA}{floating multiply-add}
\newacronym{fp}{FP}{floating point}
\newacronym{fpu}{FPU}{floating point unit}
\newacronym{ce}{CE}{compute element}
\newacronym{hwpe}{HWPE}{Hardware Processing Engine}
\newacronym{dtr}{DTR}{dual temporal redundancy}
\newacronym{hci}{HCI}{Heterogeneous Cluster Interconnect}
\newacronym{fifo}{FIFO}{first-in, first-out}
\newacronym{gemm}{GEMM}{general matrix-matrix}
\newacronym{cnn}{CNN}{convolutional neural network}
\newacronym{ci}{CI}{confidence interval}

\setcopyright{rightsretained}
\copyrightyear{2025}
\acmYear{2025}
\acmDOI{10.1145/3706594.3726981}
\acmISBN{979-8-4007-1393-4/25/05}
\acmConference[CF Companion '25]{22nd ACM International Conference on Computing Frontiers}{May 28--30, 2025}{Cagliari, Italy}
\acmBooktitle{22nd ACM International Conference on Computing Frontiers (CF Companion '25), May 28--30, 2025, Cagliari, Italy}

\begin{document}

\title{RedMulE-FT: A Reconfigurable Fault-Tolerant Matrix Multiplication Engine}







\author{Philip Wiese}
\authornote{Both authors contributed equally to this research.}
\orcid{0009-0001-7214-2150}
\affiliation{\institution{ETH Zurich}
\city{Zurich}
\country{Switzerland}}
\email{wiesep@iis.ee.ethz.ch}

\author{Maurus Item}
\authornotemark[1]
\orcid{0009-0009-6871-0272}
\affiliation{\institution{ETH Zurich}
\city{Zurich}
\country{Switzerland}}
\email{itemm@student.ethz.ch}

\author{Luca Bertaccini}
\orcid{0000-0002-3011-6368}
\email{lbertaccini@iis.ee.ethz.ch}
\affiliation{\institution{ETH Zurich}
\city{Zurich}
\country{Switzerland}}

\author{Yvan Tortorella}
\orcid{0000-0001-8248-5731}
\email{yvan.tortorella@unibo.it}
\affiliation{\institution{University of Bologna}
\city{Bologna}
\country{Italy}}

\author{Angelo Garofalo}
\orcid{0000-0002-7495-6895}
\email{angelo.garofalo@unibo.it}
\affiliation{\institution{University of Bologna}
\city{Bologna}
\country{Italy}}

\author{Luca Benini}
\orcid{0000-0001-8068-3806}
\email{lbenini@iis.ee.ethz.ch}
\affiliation{\institution{ETH Zurich}
\city{Zurich}
\country{Switzerland}}
\affiliation{\institution{University of Bologna}
\city{Bologna}
\country{Italy}}

\begin{abstract}
As safety-critical applications increasingly rely on data-parallel floating-point computations, there is an increasing need for flexible and configurable fault tolerance in parallel floating-point accelerators such as tensor engines.
While replication-based methods ensure reliability but incur high area and power costs, error correction codes lack the flexibility to trade off robustness against performance.
This work presents \RedMulE{}, a runtime-configurable fault-tolerant extension of the RedMulE matrix multiplication accelerator, balancing fault tolerance, area overhead, and performance impacts.  
The fault tolerance mode is configured in a shadowed context register file before task execution.
By combining replication with error-detecting codes to protect the data path, \RedMulE{} achieves an \resultvulnerability{} uncorrected fault reduction with only \resultareaoverheaddata{} area overhead.
Full protection extends to control signals, resulting in no functional errors after \R{\si{1}M} injections during our extensive fault injection simulation campaign, with a total area overhead of \resultareaoverheadtotal{} while maintaining a \resultfrequency{} frequency in a \SI{12}{\nano\meter} technology.
\end{abstract}



\maketitle


\section{Introduction and Related Work} \label{sec:10_introduction}
Many safety-critical systems, such as those in automotive, robotics, and space, must balance fault tolerance and computational performance to ensure reliability and efficiency~\cite{aalund_enhancing_2025}.
For instance, neural networks are frequently used in autonomous systems for sensor data processing and feature extraction~\cite{furano_ai_2020}. 
While these computations benefit from high performance, they do not necessarily require strict fault-tolerance guarantees.
In contrast, safety-critical control tasks require reliable execution to prevent mission failures. 
This increasing demand for configurable fault tolerance has led to the development of mixed-criticality systems, which integrate high-performance computing with selective fault-tolerant features, enabling a dynamic trade-off between robustness and efficiency~\cite{burns_survey_2017}.

Among the most common fault tolerance approaches, radiation-hardening by design involves techniques at the physical layout, circuit architecture, and system level to mitigate the effects of radiation-induced faults in integrated circuits. 
These approaches typically rely on physical replication or invariant-based methods to ensure computational reliability against \cglspl{see}.
While architectural replication techniques like \cgls{tmr} or \acrlong{dmr} provide strong error robustness, they incur significant area overhead.
Alternatively, invariant-based methods often do not generalize for all workloads, come with a high area overhead, or introduce substantial implementation complexity when applied to computational data paths~\cite{hamming_error_1950, lo_reliable_1994}.
Thus, the challenge in making an existing design fault-tolerant lies in balancing these countermeasures, as replication increases area significantly, while suitable invariant-based methods require deep design modifications, limiting their scalability.






To address these challenges, we propose \RedMulE{}\footnote{\R{RedMulE-FT GitHub: \repolink{}}}, a runtime-configurable fault-tolerant extension of the open-source RedMulE\footnote{RedMulE GitHub: \url{https://github.com/pulp-platform/redmule}} accelerator~\cite{tortorella_redmule_2023}.
Combining architectural replication with error-detect-ing codes, we achieve high fault tolerance while maintaining area efficiency and enabling high computational throughput.
Our approach protects the entire accelerator data and control path.
We introduce a runtime-configurable execution mechanism, allowing users to switch between a fault-tolerant mode for increased reliability and a high-performance mode for maximum computational efficiency.
The key contributions of this paper are as follows:
\begin{itemize}[nosep]
    \item We design a runtime-configurable fault-tolerant mode to protect the RedMulE accelerator against \cgls{see} and perform extensive fault simulations to characterize its robustness.
    \item The data path protection provides \resultvulnerability{} higher fault tolerance and just \resultareaoverheaddata{} area overhead over the unprotected design.
    \item With a total area overhead of \resultareaoverheadtotal{}, we achieve a fully protected configuration with no functional errors after \R{1}M injections during our fault injection simulation.
    \item Finally, we integrate \RedMulE{} into a low-power RISC-V multi-core cluster with a shared \cgls{tcdm} and implement it in a \SI{12}{\nano\meter} technology, demonstrating its feasibility in a realistic computing environment. This physical implementation demonstrates no degradation of the operating frequency (\resultfrequency{}).
\end{itemize}

While previous work has explored applying \cgls{tmr} selectively to accelerators, primarily focusing on \nobreak \acrlong{cnn}s~\cite{bertoa_fault-tolerant_2023}, our approach targets a flexible acceleration engine capable of supporting a broader range of operations. 
Moreover, a previous fault-tolerant implementation of RedMulE was proposed in~\cite{ulbricht_pulp_2023}, but it was limited to RTL-level modifications and did not include a vulnerability analysis.
Their approach uses localized checkers to protect individual \cglspl{ce}.
However, this does not protect from faults in other components, such as buffers, weight broadcast paths, and control logic.
In contrast, our work protects both the data and control path and is validated through extensive fault injection simulations.
We leverage different protection mechanisms, such as duplication, error-detecting codes, and control path redundancy to ensure system-wide fault tolerance.



\section{Background} \label{sec:15_background}

\subsection{RedMulE}
RedMulE~\cite{tortorella_redmule_2023} is a parametric, low-power accelerator for multi-precision floating-point matrix-matrix operations $\mathbf{Z} = \mathbf{Y} + (\mathbf{X} \times \mathbf{W})$, designed for integration within a PULP-based heterogeneous system. 
At its core, RedMulE features a two-dimensional array of \cglspl{ce}, where each \cgls{ce} includes a \cgls{fma} unit. 
The execution model of the accelerator optimizes data reuse by processing matrix operations row-wise while broadcasting weights column-wise to all \cglspl{ce}. 
Each row of \cglspl{ce} computes a subset of output matrix elements, accumulating results across multiple cycles under the control of a dedicated scheduler. 
RedMulE is highly configurable, with key parameters including the number of rows ($L$), the number of columns ($H$), and the number of pipeline registers per \cgls{ce} ($P$) and either FP16 or hybrid FP8 formats support. 

\subsection{PULP Cluster}
The Parallel Ultra-Low Power (PULP) cluster\footnote{\href{https://github.com/pulp-platform/pulp_cluster}{PULP Cluster GitHub: https://github.com/pulp-platform/pulp\_cluster}} is an open-source RISC-V-based computing platform for high energy efficiency. 
It features eight 32-bit RISC-V cores. 
The cores and accelerators share a \cgls{tcdm}, accessible via a single-cycle latency logarithmic interconnect. 
A dedicated \gls{dma} engine manages data transfers between \cgls{tcdm} and higher-level memory, while an AXI interface connects the PULP cluster with external subsystems.



\section{Architecture} \label{sec:20_architecture}

\RedMulE{} extends the open-source RedMulE accelerator, enabling fault tolerance through a combination of data path redundancy, control path duplication, and data protection using parity bits and \cgls{ecc}.
The primary goal of this work is to enable runtime-configurable reliability while minimizing area overhead and performance impact.
Additionally, we integrate \RedMulE{} into an enhanced version of the PULP cluster with an \cgls{ecc}-protected interconnect and \cgls{tcdm} to demonstrate feasibility in a realistic computing environment.

\begin{figure}[t]
    \centering
    \includegraphics[width=0.9\columnwidth]{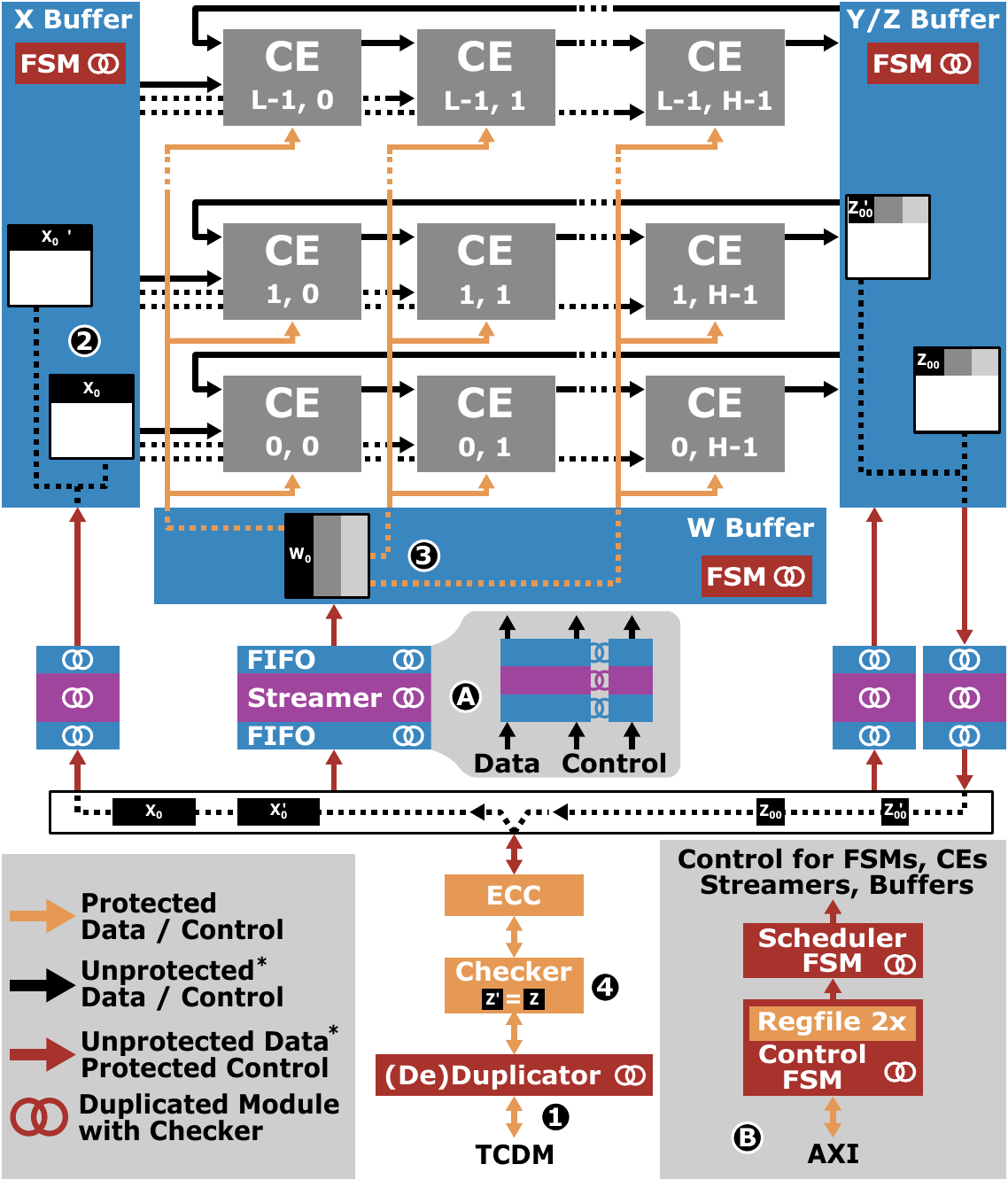}
    \setlength{\belowcaptionskip}{-15pt}
    \setlength{\abovecaptionskip}{3pt}
    
\raggedright
\footnotesize{$^*$Unprotected data links are redundant through either temporal replication or spatial duplication across consecutive compute rows.}

\caption{Architecture of \RedMulE{} with fault-tolerant data and control paths.  
\icon{1}~Duplicated read requests; filtered duplicate writes.  
\icon{2}~Redundant computation on consecutive rows.  
\icon{3}~Parity-protected broadcasted weights.  
\icon{4}~Final results checked for equality.  
\icon{A}~Duplicated modules with reduced data width for control protection.  
\icon{B}~Duplicated FSMs with parity-protected register file.}
\label{fig:20_architecture_redmule}
\Description[Architecture of \RedMulE{} with fault-tolerant data and control paths.]{
Architecture of \RedMulE{} with fault-tolerant data and control paths.  
\icon{1}~Duplicated read requests; filtered duplicate writes.  
\icon{2}~Redundant computation on consecutive rows.  
\icon{3}~Parity-protected broadcasted weights.  
\icon{4}~Final results checked for equality.  
\icon{A}~Duplicated modules with reduced data width for control protection.  
\icon{B}~Duplicated FSMs with parity-protected register file.
}
\end{figure}

\subsection{Data Path Redundancy} 
\label{sec:20_architecture:data_path_redundancy}
Unlike \cite{ulbricht_pulp_2023}, we do not introduce one checker per \cgls{ce}, as it only provides localized protection.
Instead, \RedMulE{} leverages its internal parallelism to achieve fault tolerance with minimal area overhead, as shown in \Cref{fig:20_architecture_redmule}, while overlapping protection mechanisms to enhance fault tolerance at the system level.
Since the accelerator performs \cglspl{fma} across multiple parallel rows, we introduce redundancy by ensuring that two consecutive rows compute the same result. 
Data loading in RedMulE is handled individually per lane, allowing duplicated calculations by adapting the accelerator \cgls{fsm} behavior. 
For the $\mathbf{X}$ and $\mathbf{Y}$~inputs, elements are fetched row-wise for each row of \cglspl{ce}. 
To introduce redundancy, each memory access response is duplicated before \cgls{ecc} decoding, ensuring that two consecutive rows receive identical data without generating unnecessary \cgls{tcdm} access.
This redundancy allows fault detection at the output stage using a checker mechanism that compares the duplicated results.
For the $\mathbf{Z}$ output elements, redundancy is introduced by storing two rows at a single memory address. 
A checker verifies the outputs to detect faults before storing the result.
Additionally, a hardware filter is integrated within \RedMulE{} to prevent redundant write requests to the \cgls{tcdm}.
Since the weights $\mathbf{W}$ are broadcasted to all rows, a single corrupted element would propagate to multiple computations.
Thus, we introduce parity verification at each \cgls{ce} post-broadcast, ensuring that faults in weights are detected at the point of use with minimal overhead.


\subsection{Control Path Protection}\label{sec:20_architecture:control_path_protection}

Since errors in the load-store path are detectable by duplicate loading, full replication is unnecessary.
We duplicate buffers and streamers with a reduced data width while still generating all required control signals for comparison.
This approach allows us to protect the control logic while significantly reducing the total area overhead compared to full duplication.
The redundancy mechanism for $\mathbf{W}$ inputs, as described in \Cref{sec:20_architecture:data_path_redundancy}, indirectly also protects the control signals for the $\mathbf{W}$-data path, as the parity bits are generated by separate logic. 
This separation prevents a single corrupted control signal from affecting both the $\mathbf{W}$-data and its associated parity bits simultaneously.
Any misalignment between the two indicates a fault, ensuring robust protection.
To ensure the data integrity of the configuration in the register file, we extend it with \texttt{XOR}-based parity bits computed by the cluster cores.
Since this affects only a few configuration values, the overhead is limited to a one-time increase of \R{\SI{120}{cycles}} per workload at most.
For the remaining control elements, such as the control and schedule \cglspl{fsm} and parity checker of the register file, we duplicate the respective modules and compare their outputs to detect inconsistencies. 
To further protect the connection between the control instances and \cglspl{ce}, compute rows are alternately assigned to either the primary or replica control \cgls{fsm}. 

\subsection{Fault Handling and Integration}
During operation, \RedMulE{} continuously verifies the integrity of the register file using the provided parity bits and checks the output of the different checkers.
If a fault is detected, several actions are triggered: (1) the fault status registers are updated with the detected error information, (2) an interrupt is raised to notify the host system, and (3) the \cgls{fsm} of \RedMulE{} returns into the idle state, allowing the system to start another operation or repeat the current one.
To ensure reliable fault notification to the host, the interrupt signal is asserted for two consecutive cycles. 
This mechanism guarantees that even in the presence of a transient fault on the interrupt wire, the host system receives the signal correctly.
Once the host system acknowledges the interrupt, it can read and clear the fault status registers before initiating a new computation on \RedMulE{}.

\subsection{Runtime Mode Configuration}
\RedMulE{} can operate in the fault-tolerant mode for improved reliability or performance mode for maximum throughput.
The mode selection is done before computation by configuring a register in the accelerator register file.
In fault-tolerant mode, redundant computations and error-checking mechanisms are enabled, ensuring maximum reliability.  
In performance mode, the control path redundancy is still active, but computations are not duplicated, and detected faults will lead to an abort of the current workload.



\section{Evaluation} \label{sec:30_implementation}
To assess the impact of our fault-tolerant implementation, we synthesized \RedMulE{} with $L=12$ rows, $H=4$ \cgls{ce} per row, $P=3$ pipeline registers per \cgls{ce}, and FP16 support in GlobalFoundries' 12LP+ FinFET technology. 
The design was synthesized and implemented under worst-case conditions \R{(SS, \SI{0.72}{\volt}, \SI{125}{\celsius})}, targeting an operating frequency of \resultfrequency{}.
We evaluate three versions of RedMulE to analyze the trade-offs between area and fault tolerance:
\aptLtoX[graphic=no,type=html]{\begin{enumerate}
    \item The baseline non-protected RedMulE presented in~\cite{tortorella_redmule_2023}.
    \item A partially protected \RedMulE{}, which only protects the data path as described in \Cref{sec:20_architecture:data_path_redundancy}. 
    \item The fully protected \RedMulE{}, which includes the control path redundancy as described in \Cref{sec:20_architecture:control_path_protection}.
\end{enumerate}}{\begin{enumerate}[label=(\arabic*),nosep]
    \item\label{ite:var_1} The baseline non-protected RedMulE presented in~\cite{tortorella_redmule_2023}.
    \item\label{ite:var_2} A partially protected \RedMulE{}, which only protects the data path as described in \Cref{sec:20_architecture:data_path_redundancy}. 
    \item\label{ite:var_3} The fully protected \RedMulE{}, which includes the control path redundancy as described in \Cref{sec:20_architecture:control_path_protection}.
\end{enumerate}}

\subsection{Physical Implementation}

\begin{figure}[t]
    \centering
     \begin{subfigure}[][][c]{1\columnwidth}
        \centering
        \includegraphics[width=0.85\textwidth]{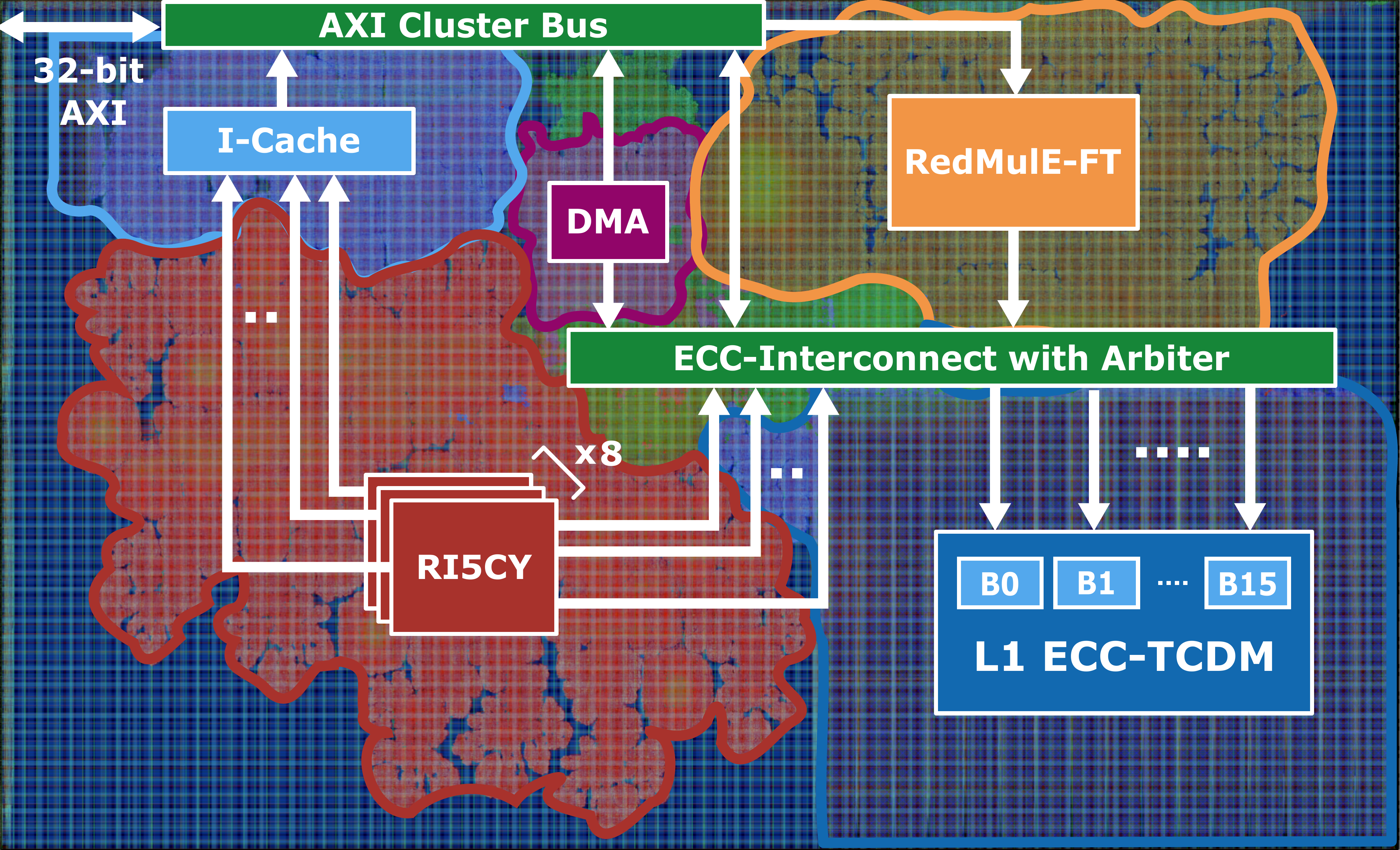}
        \caption{Architecture of the PULP cluster with \RedMulE{} inside a placed-and-routed \SI{1400}{\micro\meter}~$\times$~\SI{850}{\micro\meter} block.}
        \label{fig:30_implementation_chip_physical}
    \end{subfigure}
    \hfill
     \begin{subfigure}[][][c]{1\columnwidth}
        \centering
        \includegraphics[width=0.9\textwidth]{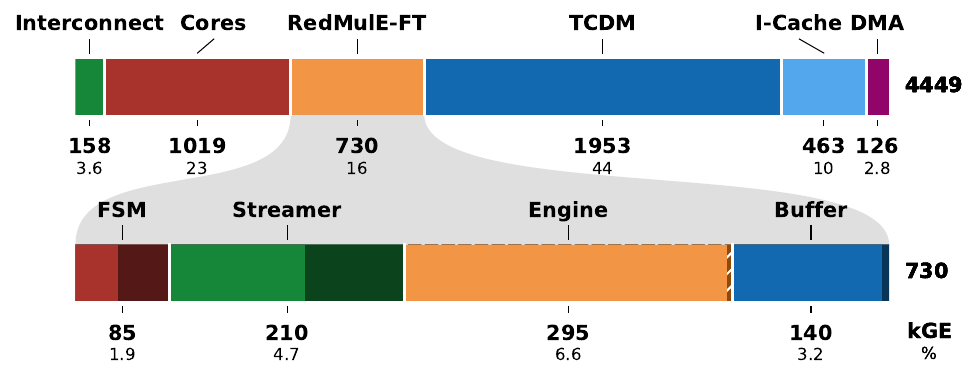}
        \caption{Area breakdown with the hatched parts indicating the area overhead compared to the baseline RedMulE.}
        \label{fig:30_implementation_chip_area}
    \end{subfigure}
    \setlength{\belowcaptionskip}{-15pt}
    \setlength{\abovecaptionskip}{3pt}
    \caption{PULP cluster with full protected \RedMulE{}, implemented in GlobalFoundries' 12LP+ FinFET.}
    \label{fig:30_implementation_chip}
    \Description[Area breakdown of the PULP cluster with fully protected \RedMulE{}]{A bar chart representing the area breakdown of RedMulE-FT and its surrounding components
within the PULP cluster. Each section is labeled with its corresponding
percentage of total area usage, including Interconnect (3.6 \%),
Cores (23 \%), RedMulE-FT (16 \%), TCDM (44 \%), I-Cache (10 \%), and
DMA (2.8 \%). Additionally, subcomponents of RedMulE-FT are detailed,
such as FSM (1.9 \%), Streamer (4.2 \%), Engine (6.7 \%), and Buffer
(3.2 \%).}
\end{figure}

We implemented the complete PULP cluster to verify the physical implementation, as shown in \Cref{fig:30_implementation_chip}a\footnote{Rendered with Artistic:  \url{https://github.com/pulp-platform/artistic}}.
While the baseline RedMulE accelerator \ref{ite:var_1} occupies \resultareabaseline{}, adding data protection to RedMulE \ref{ite:var_2} introduces a \resultareaoverheaddata{} increase in area, resulting in \resultareadata{}. 
This overhead is primarily attributed to \cgls{ecc} encoders/decoders and the data checkers within the streamer, together with \cgls{ecc} and data faults tracking registers and more complex address generators.

Extending fault protection to control signals in \ref{ite:var_3} introduces an additional area increase of \resultareaoverheadfull{}, bringing the total overhead for full redundancy to \resultareaoverheadtotal{}, resulting in \resultareafull{}, as shown in \Cref{fig:30_implementation_chip}b.
This increase is primarily due to the duplication of the streamer modules, control, and scheduler \cglspl{fsm}.
While the overhead appears large, it reflects the small compute data path of this RedMulE instance, where control structures form a significant portion of the total area.
The relative cost of fault tolerance would considerably decrease in larger configurations with more \cgls{fma} units.
Despite these modifications, the fault-tolerant configurations \ref{ite:var_2} and \ref{ite:var_3} do not affect the critical path, thus maintaining the same operating frequency of \R{\SI{500}{\MHz}} as the baseline design \ref{ite:var_1}. 
Both configurations support a runtime-configurable fault-tolerant mode, which can be switched off to disable the redundancy and compute in an unprotected execution at $2\times$ the performance.
If a fault is detected in redundant mode, the computation is terminated, the accelerator is re-programmed, and a full re-execution is initiated in fault-tolerant mode, further increasing execution time.
However, as the number of retries remains relatively low in practice, the overall performance impact is manageable, as demonstrated by the fault-injection analysis in the following subsection.

\subsection{Fault-Injection Analysis}
To evaluate the robustness of \RedMulE{}, we performed extensive fault injection simulations.
In the simulations, single faults were injected into combinational nets of different RedMulE versions while executing a matrix multiplication workload with dimensions $(12 \times 16 \times 16)$.
The clock tree and reset network were excluded from the fault injection, and we assume that no additional faults occur during the recomputation phase triggered by the fault recovery mechanism.
Error bounds are computed using a Poisson distribution with a \SI{95}{\percent} \cgls{ci} and conservatively assuming one additional observed error.








\begin{table}[t]
\centering
\setlength{\belowcaptionskip}{-5pt}
\setlength{\abovecaptionskip}{5pt}
    \caption{Fault injection results for different RedMulE versions, based on \R{\si{1}M} injected faults per configuration.}
    \label{tab:40_redmule_vulnerability}
    \begin{tabular}{@{}l@{\hspace{0.5em}}*2S[table-format=2.2,table-space-text-post=\%]S[table-format=3.4,table-space-text-post=\%]@{}}\toprule
        & \textbf{Baseline}  & \textbf{Data} & \textbf{Full}~\textsuperscript{a)} \\
        & \textbf{\cite{tortorella_redmule_2023}} & \textbf{Protection} & \textbf{Protection} \\
        \midrule
        \textbf{Correct Termination} & \hphantom{00}\textbf{92.92\,\%} & \textbf{99.36\,\%} & \textbf{>99.9997\,\%} \\
        \quad w/o Retry & 92.92\,\% & 88.01\,\% & 87.4457\,\% \\
        \quad with Retry & 0.00\,\% & 11.35\,\% & 12.5543\,\% \\
        \textbf{Functional Error} &  \hphantom{00.}\textbf{7.08\,\%} &  \textbf{0.65\,\%} &  \hphantom{.}\textbf{<0.0003\,\%} \\
        \quad Incorrect & 6.97\,\% & 0.46\,\% & <0.0003\,\% \\
        \quad Timeout & 0.11\,\% & 0.19\,\% & <0.0003\,\% \\
        \hline
        \textbf{Area Overhead} &  \hphantom{00.}\textbf{0.0\,\%} &   \textbf{\resultareaoverheaddata{}} &  \textbf{\resultareaoverheadtotal{}} \\
        \bottomrule
    \end{tabular}
    \raggedright
    \footnotesize\textsuperscript{a)}~No incorrect results and timeouts observed. Upper bound estimated via Poisson distribution (\SI{95}{\percent} \gls{ci}), assuming one additional error.
\vspace{-15pt}
\end{table}



In the baseline configuration \ref{ite:var_1}, only \R{\SI{7.08(0.05)}{\percent}} of the injections lead to functional errors, as shown in Table~\ref{tab:40_redmule_vulnerability} (Column 1).
This relatively low vulnerability can be attributed to the fact that not all parts of the circuit are actively used at any given time. 
If a transient fault occurs in an idle section, it does not necessarily propagate to the output, thereby mitigating the overall error rate.
Introducing data protection in \ref{ite:var_2} significantly improves fault tolerance. 
Functional errors are significantly reduced, resulting in \R{\SI{99.36(0.02)}{\percent}} correct terminations.
Compared to the baseline \ref{ite:var_1}, most faults that previously resulted in functional errors now result in correct termination after recomputation.
Consequently, the data protection of \RedMulE{} reduces the vulnerability by \resultvulnerability{} compared to the baseline implementation. 
With the fully protected version of \RedMulE{} \ref{ite:var_3}, more than \R{\SI{99.9997}{\percent}} of injected faults result in correct termination with no functional errors after \R{\si{1}M} injections.
Approximately \R{\SI{12}{\percent}} of injected faults require a recomputation, indicating that the retry mechanism effectively handles detected errors by restarting operations when necessary.
Additionally, no timeouts were observed, suggesting that the fault recovery mechanism successfully mitigates control path corruption.



\section{Conclusion} \label{sec:50_conclusion}
In this work, we introduced \RedMulE{}, a runtime-configurable fault-tolerant accelerator, and evaluated its vulnerability under fault injection. 
By leveraging parallelism and introducing redundancy with parity-protected data, we achieved an \resultvulnerability{} improvement in fault tolerance with only \resultareaoverheaddata{} area overhead.
Extending protection to the control path resulted in no functional errors after \si{1}M injections, ensuring reliable execution after a retry, with \resultareaoverheadtotal{} area overhead.
While fault-tolerant mode reduces performance by $2\times$ due to duplicated computation, it ensures robust error detection and correction.
In contrast, performance mode fully utilizes all compute resources but does not detect errors, making it suitable for non-critical workloads.
Future work could refine fault recovery to prevent full matrix recomputation, enabling tile-level recovery with a more sophisticated resynchronization mechanism.


\begin{acks}
This work was supported in part through the TRISTAN (101095947) project that received funding from the HORIZON CHIPS-JU programme.
It also received funding from the Swiss State Secretariat for Education, Research, and Innovation (SERI) under the SwissChips initiative.
\end{acks}

\bibliographystyle{ACM-Reference-Format}




\end{document}